\documentclass[twocolumn,prl,showpacs]{revtex4}

\usepackage{graphicx}
\usepackage{rotating}
\usepackage{amsmath}
\usepackage{amsfonts}
\usepackage{amssymb}
\usepackage{enumerate}
\usepackage{longtable}
\usepackage{dcolumn}
\usepackage{bm}

\begin{document}

\newcommand{\be}{\begin{equation}}
\newcommand{\ee}{\end{equation}}
\newcommand{\bn}{\begin{eqnarray}}
\newcommand{\en}{\end{eqnarray}}

\title{Theory of Orbital Nematicity in Underdoped Iron Arsenides}

\author{M. S. Laad$^1$ and L. Craco$^2$}

\affiliation{$^1$Institut f\"ur Theoretische Physik, RWTH Aachen University,
52056 Aachen, Germany \\ 
$^2$Physical Chemistry, Technical University Dresden,
01062 Dresden, Germany}

\date{\today}

\begin{abstract}
Recent finding of an {\it unusual} in-plane resistivity anisotropy in the 
underdoped 122-family at high temperature ($T$) suggests an orbital nematic 
(ON) order, posing a challenge to extant theories.  The {\it sign} of the 
anisotropy contradicts expectations from weakly correlated as well as pure 
spin-only nematic views. Here, we show how such an ON order with accompanying 
structural distortion arises from {\it residual}, intersite and 
inter-orbital two-body interactions in an incoherent ``bad metal'' close to 
Mottness. Enhancement of orbital-selective incoherence is shown to be 
necessary for understanding transport anisotropy. Our results suggest that 
ON order, with subsequent antiferromagnetic order might be the {\it primary}
competitor to superconductivity in Fe arsenides.     
\end{abstract}


\maketitle

The ``Iron Age'' of high temperature superconductivity (HTSC) appears to 
reinforce the paradigm shift set in motion by HTSC in cuprates more than 
twenty years ago. Examining analogies and differences between cuprates 
and Fe-pnictides (FePn) can shed more light on both.  While parent cuprates are 
antiferromagnetic (AF) Mott insulators, parent FePn are multi-band 
AF metals, with the Fermi surface (FS) constituted of several small  
pockets.  Notwithstanding this difference, the normal metallic state 
(by chemical doping) in both defies even a qualitative description within 
traditional Landau Fermi Liquid (LFL) concepts.  FePn thus 
appear to be anomalous metals giving way to unconventional superconductivity 
at high temperature ($T$)~\cite{louis}.

AF in FePn is universally preceded by, or coincides with, a 
tetragonal-to-orthorhombic (T-O) structural distortion. Recent work on
122-~\cite{fisher} and 1111-based~\cite{buechner} FePn reveals 
signs of enhanced nematic (N) susceptibility ($\chi_{N}$) and {\it electronic} 
crystallinity upon underdoping, deepening the analogy with cuprates. Transport 
data puts strong constraints on theories purporting to describe the N-phase. 
In particular~\cite{fisher}, (i) inspite of the fact that the N order leads 
to $b<a$ (here, $a,b$ are planar unit cell distances), and (ii) the 
${\bf q}=(\pi,0)$ AF implies antiferromagnetically aligned spins along $a$ 
and ferromagnetically aligned spins along $b$, the resistivity $\rho_{b}(T)$ 
along $b$ {\it exceeds} $\rho_{a}(T)$ along $a$, an inexplicable result in a 
weak-coupling LFL~\cite{mimin} or pure spin-fluctuation~\cite{j1j2} scenarios. 
In addition, $\rho_{b}(T)$ shows insulator-like behavior above $T_{c}$, even 
as $\rho_{a}(T)$ remains metallic. These observations call for a more careful 
investigation into microscopic sources of nematicity, especially in view of 
the resistivity anisotropy peaking (maximal nematic response, $\chi_{N}$) 
around a doping where SC $T_{c}(x)$ is highest.  Thus, the incoherent 
``normal'' state gives way to either (a) a nematic phase (at least in 
122-FePn), with a T-O distortion and subsequent SDW order, upon underdoping, 
or (b) to an unconventional SC with a dome in $T_{c}$ versus $x$, or (c) to 
a renormalized Fermi liquid metal at low $T$ upon overdoping.
        
These findings in 122-FePn conflict with {\it pure} spin 
fluctuation models ((i) and (ii) above), and, in addition to (i),(ii), 
in view of the insulator-like $\rho_{b}(T)$, weakly correlated models as 
well. Orbital-based models do identify the T-O distortion with multi-orbital 
($xz,yz$) correlations~\cite{phillips}, but nematicity has hitherto not been 
considered there.  Finally, how such a coupled nematic-plus 
T-O transition could arise as an instability of the ``high''-$T$ ``strange'' 
metal~\cite{louis,hess} remains open (the T-O transition generically precedes 
AF-spin density wave (SDW)). Here, we address these issues in detail using 
LDA+DMFT. Our philosophy is to identify the germs of the emergent order in 
the residual (two-particle) interactions present in the high-$T$ incoherent 
fluid. Thereupon, extending our earlier LDA+DMFT calculations to a 
postulated orbital-nematic (ON) state, we show how the unusual features above 
find natural explication within a theoretical picture where an orbital 
nematic order arises as a direct instability of a sizably correlated 
bad metal.

In an incoherent metal, inter-band one-electron mixing is {\it irrelevant} 
in a renormalization-group (RG) sense, akin to the situation that occurs 
in weakly coupled $D=1$ Luttinger liquids~\cite{pwa}. This inhibits a smooth 
crossover to a correlated FL, simultaneously favoring ordered states arising 
directly from the incoherent metal. The underlying reason is that when 
one-electron mixing is irrelevant, the corresponding second-order process, 
corresponding to two-particle intersite {\it and} inter-orbital hoppings, 
turns out to be the most relevant.  In our earlier LDA+DMFT for the 
``normal'' state, an incoherent metal arises from an Anderson orthogonality 
catastrophe-like effect due to orbital-selective (OS) Mott physics. The general
 two-body interaction, obtained at second order from a (now RG-irrelevant) 
one-electron inter-band term, 
$t_{ab}\sum_{<i,j>,\sigma}(c_{ia\sigma}^{\dag}c_{jb\sigma}+h.c)$, is 
$H^{(2)}=\frac{1}{2}\sum_{a,b,k,k'}V_{ab}(k,k') c_{a,k,\uparrow}^{\dag} 
c_{b,-k,\downarrow}^{\dag}c_{b,-k',\downarrow}c_{a,k',\uparrow}$,
where $a,b=xz,yz$ and the scattering vertex is
$V_{ab}(k,k',\omega)=g^{2}\chi_{ab}(k-k',\omega)$ with $\chi_{ab}(k,\omega)$ 
being the inter-orbital susceptibility.  The static, nearest- and 
next-nearest neighbor parts of 
$V_{ab}(k,k')$ are $V_{ab}^{(1)}(k,k',0)\simeq \frac{4t_{ab}^{2}}{U'+J_{H}}\simeq
O(200-250)$~meV and $V_{ab}^{(2)}\simeq \frac{4t_{ab}'^{2}}{U'+J_{H}}\simeq
O(120-160)$~meV, consistent with magnetic excitation energy scales from 
INS data. These {\it residual} interactions contain coupled, inter-orbital 
charge and spin fluctuations in particle-hole (p-h) {\it and} cooper (p-p)
channels, immediately suggesting the possibility of both, mutually competing 
instabilities~\cite{sachdev} from the ``high'' $T$ incoherent metal. 
In earlier work, we showed how an unconventional SC with the right gap 
symmetry indeed arises from the cooper instability above~\cite{laad}. Here, 
we study how a p-h decoupling of $H^{(2)}$ above in the situation specific 
to FePn leads to the ON instability.  We decouple $H^{(2)}$ in the 
p-h sector as 
$H_{MF}^{(2)}=\sum_{a,b,k,\sigma}(\Delta_{ab}^{(n)}(k,k')c_{a,k,\sigma}^{\dag} 
c_{b,k',\sigma}+a\rightarrow b)$, 
with $\Delta_{ab}^{(n)}(k,k')=\frac{1}{2}V_{ab}(k,k',0)\langle
c_{a,k,\sigma}^{\dag}c_{b,k',\sigma}\rangle$. 
Consistent with lattice symmetries,
we write $V_{ab}(k,k')=\sum_{l}V_{ab}^{l}\eta_{l}(k)\eta_{l}(k')$, with 
$\eta_{l}(k)$ being the irreducible representations of the $D_{4h}$ point 
group.  Then it follows that 
$\Delta_{ab}(k\simeq k')=\sum_{l}\Delta_{ab}^{l}\eta_{l}(k)$, and, for the 
frustrated geometry of FePn,
$\Delta_{ab}(k)=\Delta_{ab}^{(1)}(c_{x}+c_{y})+\Delta_{ab}^{(2)}c_{x}c_{y}$ 
where $c_{x}=$cos$k_{x}$, etc.

Interestingly, $N^{z}\simeq\langle n_{xz}-n_{yz}\rangle$ with 
$\langle N_{a}\rangle=\langle\sum_{k}\Delta_{ab}^{(n)}(k)c_{a,k,\sigma}^{\dag}c_{a,k,\sigma}\rangle$ (with $a=xz,yz$), 
which is now finite, is exactly associated with
{\it orbital} nematicity. 
And $\alpha=\Delta_{ab}^{(2)}/\Delta_{ab}^{(1)}\simeq 0.7$. 
A $\langle N\rangle=\frac{\langle n_{xz}\rangle-\langle n_{yz}\rangle}{\langle
n_{xz}\rangle+\langle n_{yz}\rangle}\ne 0$ lifts the degeneracy of the 
$xz,yz$ bands in the T-structure, as seen by considering that: 
(i) its form factor, $\Delta_{ab}^{(n)}(k)$, differs from those for the 
$xz,yz$ bands (ii) in the real FePn structure, details of $p-d$ orbital 
overlaps along $a$ and $b$ axes leads to important differences in the 
$xz,yz$ band dispersion~\cite{belen} already at LDA and Hartree-Fock (HF) 
levels. With an anisotropic form factor in (i) above, we get
$\langle N_{e}\rangle >0$~\cite{wu}, as in bilayer ruthenates, already at HF 
level.  Note that this is now an ON instability arising from 
{\it an interplay between residual interactions} and {\it the renormalized 
LDA+DMFT band structure}, and arises directly from the incoherent metal. 
Our mechanism is thus radically different from one where ordered states
arise from a weakly correlated LFL metal~\cite{mimin}.  A finite 
$\Delta_{ab}^{(n)}(k)$ will split the $xz,yz$ degeneracy above  and modify 
the $xz,yz$ band {\it dispersion} anisotropically, with an increase in 
$xz$ orbital character in the ``reconstructed'' FS across the T-O transition, as noted recently~\cite{belen}, an effect we will consider in future work. 
The {\it symmetry} dictated coupling to the sizable magneto-elastic 
interaction in FePn will reinforce this splitting by coupling 
$\langle N \rangle$ linearly to the strain.  
This acts like a sizable ``external'' 
field, potentially enhances
extant nematic order and will smear the nematic transition, making 
it impossible to separate nematicity from the T-O distortion.

If the $xz$ band is thereby pushed below its position in the T-phase, it 
is not hard to deduce (see results below) that $n_{xz}>n_{yz}$, giving a 
finite value for $\langle N \rangle$. 
Strong local correlations ($U,U',J_{H}$) will now have a stronger 
(``Mott'') localizing effect on the (more populated) $xz$ band states, 
and we propose that the shortening of the lattice constant $b$ is thus 
a direct fall-out of this correlation-induced instability to the 
ON phase.  This will immediately induce the T-O distortion ($b<a$) 
and create conditions propitious for a ${\bf q}=(\pi,0)$ AF instability. 
Enhancement of Mottness in the $xz$ sector, together with the incoherent 
``normal'' state above $T_{c}$, is also consistent with the observation 
of an insulator-like $\rho_{b}(T)$ below $200$~K in the 122-systems, 
and cannot be found in weak-coupling pictures. Thus, the observed ON-phase 
(competing with U-SC) is intimately linked to {\it enhancement} of the OS 
incoherence, already a salient feature of the ``normal'' state.  Sizable 
multi-orbital (MO) correlations and OS Mottness thus turn out to be 
{\it necessary} for even a qualitative reconciliation of these unusual 
features, as we detail below. 

Extending earlier LDA+DMFT studies, we focus on a {\it four band} 
model, since the ON phase involves predominantly planar bands.  Defining 
$\delta=\frac{b-a}{b+a}$ as the orthorhombicity, the orbital-lattice coupling 
is $H_{e,l}=\lambda\delta(n_{xz}-n_{yz}) + \frac{C_{s0}}{2}\delta^{2}$, 
{\it a la} the more familiar Jahn-Teller coupling in MO systems. 
This self-consistently renormalizes {\it both}, $\langle N\rangle$ 
and $\delta$, as alluded to before. Since {\it ab initio} values of 
$\lambda,C_{s0}$ are not known, we adopt the strategy of using trial values 
for $\langle N'\rangle=\langle n_{xz}-n_{yz}\rangle_{0}=-\frac{C_{s0}\delta}{\lambda}$ in addition to the $\langle N_{e}\rangle$ computed from 
$\langle\sum_{k,\sigma}(n_{k,a,\sigma}-n_{k,b,\sigma})\rangle$ computed within 
LDA+DMFT~\cite{wu}. We input trial values of 
$\langle N\rangle=\langle N_{e}\rangle+\langle N'\rangle$
as ``external'' symmetry breaking fields in the $xz,yz$ orbital sector, 
and recompute the LDA+DMFT spectral functions in the sizably correlated 
multi-band Hubbard model~\cite{1111-prb}, but now including the finite 
$\langle N\rangle$ above, along with $U=4.0$~eV, $U'=2.6$~eV and 
$J_{H}=0.7$~eV, as used in earlier work. $\langle N^{z}\rangle$ is 
now self-consistently computed within DMFT from the converged 
$G_{xz}(k,\omega),G_{yz}(k,\omega)$.

\begin{figure}
\includegraphics[width=3.8in]{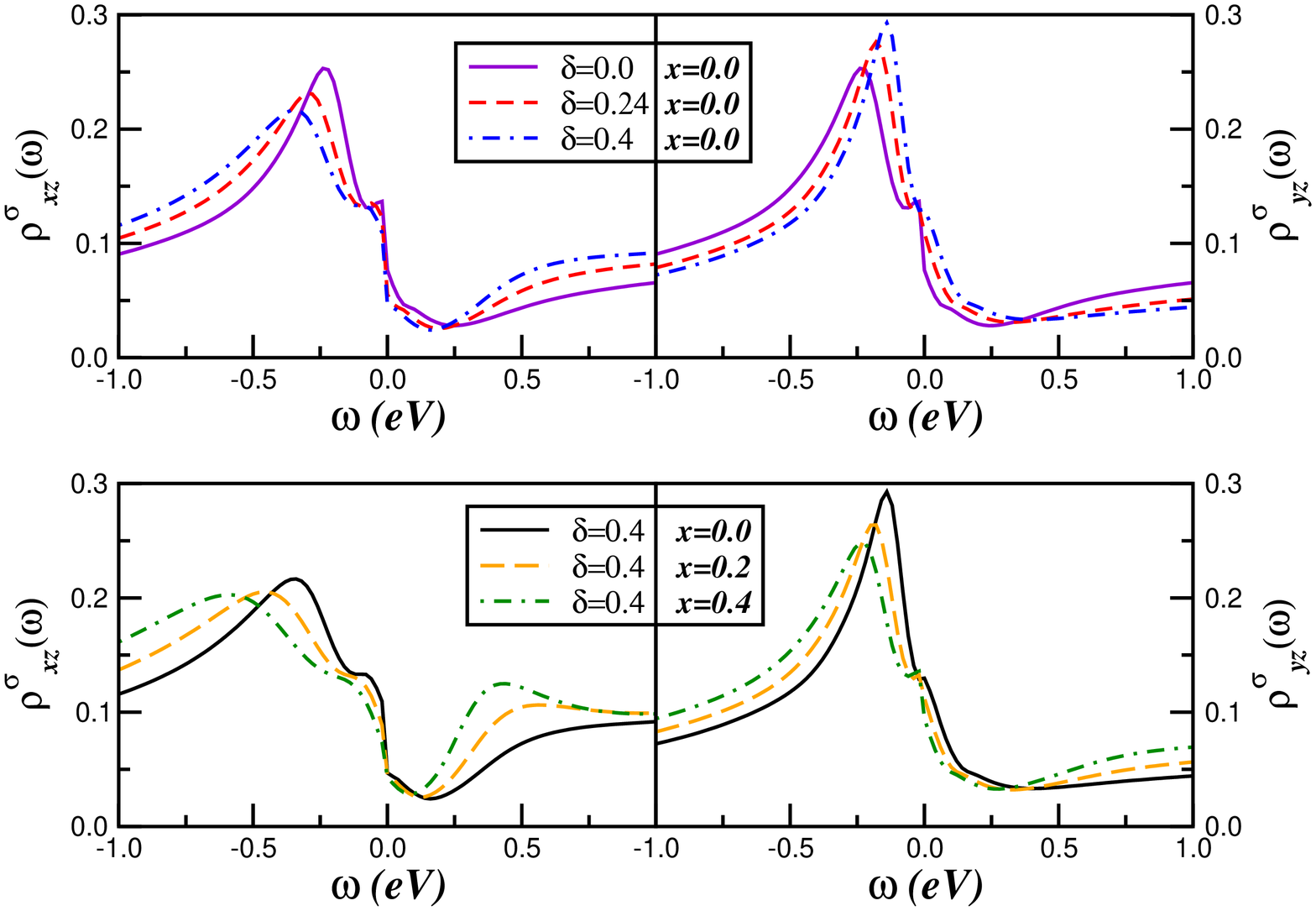}
\caption{(Color online)
Orbital-resolved LDA+DMFT (with $U=4.0$~eV, $U'=2.6$~eV and
$J_{H}=0.7$~eV) density-of-states (DOS) for the Fe $d$ orbitals for
three $\delta$ values (top) and three doping values with fixed $\delta=0.4$ 
(bottom).  Large-scale and {\it anisotropic} dynamical spectral weight transfer 
accompanying the incoherent metal-nematic metal transition is clear, in good accord with optics~\cite{fisher1}.}
\label{fig1}
\end{figure}

\begin{figure}[t]
\includegraphics[width=3.8in]{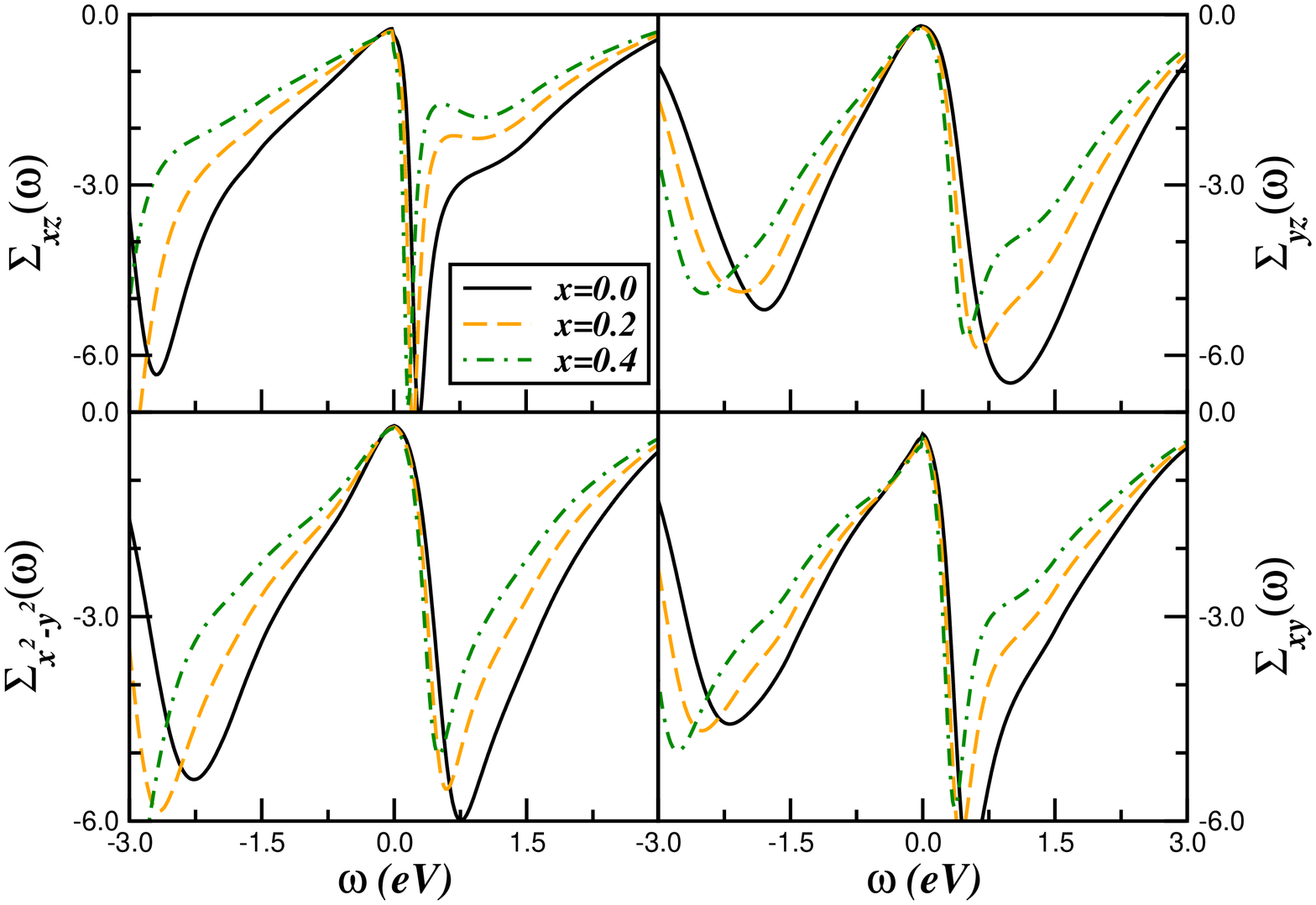}
\caption{(Color online)
Imaginary parts of the orbital-resolved LDA+DMFT self-energies 
for $\delta=0.4$, showing sub-linear in-$\omega$ dependence for
$\omega$ close to $E_{F}$, reflecting an incoherent metal. Also, 
the sharp pole-like feature in Im$\Sigma_{xz}$ progressively 
sharpens and moves closer to $E_F$ with $\delta$, reflecting enhanced OS 
incoherence.}
\label{fig2}
\end{figure}

We now discuss our results.  In Fig.~\ref{fig1}, we show the (converged)
LDA+DMFT spectral functions for the in-plane $d$ orbitals with 
$\langle N\rangle=0$ and $\langle N\rangle\ne 0$.  Clear low-energy OS 
incoherence is seen in the ``normal'' metallic (T) phase, a direct 
consequence of sizable MO correlations, as noticed by many workers~\cite{1111-prb}. 
With $\langle N\rangle\ne 0$, closer examination reveals interesting 
changes germane to the above discussion: (i) $\langle N\rangle >0$ and 
``zeeman field'' ($h_{z}>0$) remove the $xz-yz$ degeneracy of the 
$T$-phase. With $U=4.0$~eV, $U'=2.6$~eV, MO Hartree shifts from sizable 
MO correlations renormalize $h_{z}=\delta$.  DMFT values for 
$\langle N\rangle$ are found to be 0.049 for $x=0.2 (\delta=0.4)$~eV
and 0.0315 for $x=0.4 (\delta=0.24)$~eV, vanishing in the $T$ phase.
Concomitantly, $\delta(T)$ follows the ON order parameter via
$\delta=-\frac{\lambda}{C_{s0}}\langle N\rangle$.  The dynamical 
self-energies ($\Sigma_{a}(\omega)$) from DMFT then drive sizable 
{\it changes} in spectral weight transfer (SWT) over energies 
$O(1.0-2.0)$~eV in response to the renormalization of $\delta$. This 
feature is generic to strongly correlated quantum matter: observation 
of sizable SWT across the T-O transition should put our approach on 
firmer ground. (ii) revealingly, the pseudogap (PG), already a feature 
of the ``normal'' state, deepens for the $xz$ band and {\it reduces} for 
the $yz$ band. Concomitantly, the pole-like structure in 
Im$\Sigma_{xz}(\omega)$ progressively sharpens up and moves closer to 
$E_{F}$ with increasing $x$ (Fig.~\ref{fig2}) for not-too-large $x$.  
Finally, given sizable $U'$, interband ``dynamical proximity'' effects 
induce sizable spectral changes in the $xy,x^{2}-y^{2}$ spectra as well. 
Thus, OS incoherence, already present in the ``normal'' 
state~\cite{1111-prb,vojta}, is further enhanced across the ON 
transition (iii) Using 
$n_{a}=(-1/\pi)\int_{-\infty}^{E_{F}}$Im$G_{a}(\omega)d\omega$, we find 
$n_{xz}>n_{yz}$, as advertized above.  Thus, $\langle N\rangle >0$ and, 
given the enhanced (Mott-like) localization in the $xz$ sector, the lattice 
constant $b$ now shrinks relative to its value ($b=a$) in the T-phase: 
this is precisely the T-O distortion.  With finite $\langle N\rangle\ne 0$ 
tied to enhanced Mott-like localization in the $xz$-sector, one can now 
write an {\it effective} ferro-nematic pseudospin model~\cite{phillips}: 
$H_{eff}\simeq -j\sum_{<i,j>}N_{i}^{z}N_{j}^{z}-\delta(T)\sum_{i}N_{i}^{z}$,
and, with $\delta(T)$ as chosen above, {\it structural} anisotropy 
disappears at $T_{T-O}$.  But strong precursor nematic correlations survive
appreciably above $T_{T-O}$ (cf. Ising model in a zeeman field).  

\begin{figure}
\includegraphics[width=3.6in]{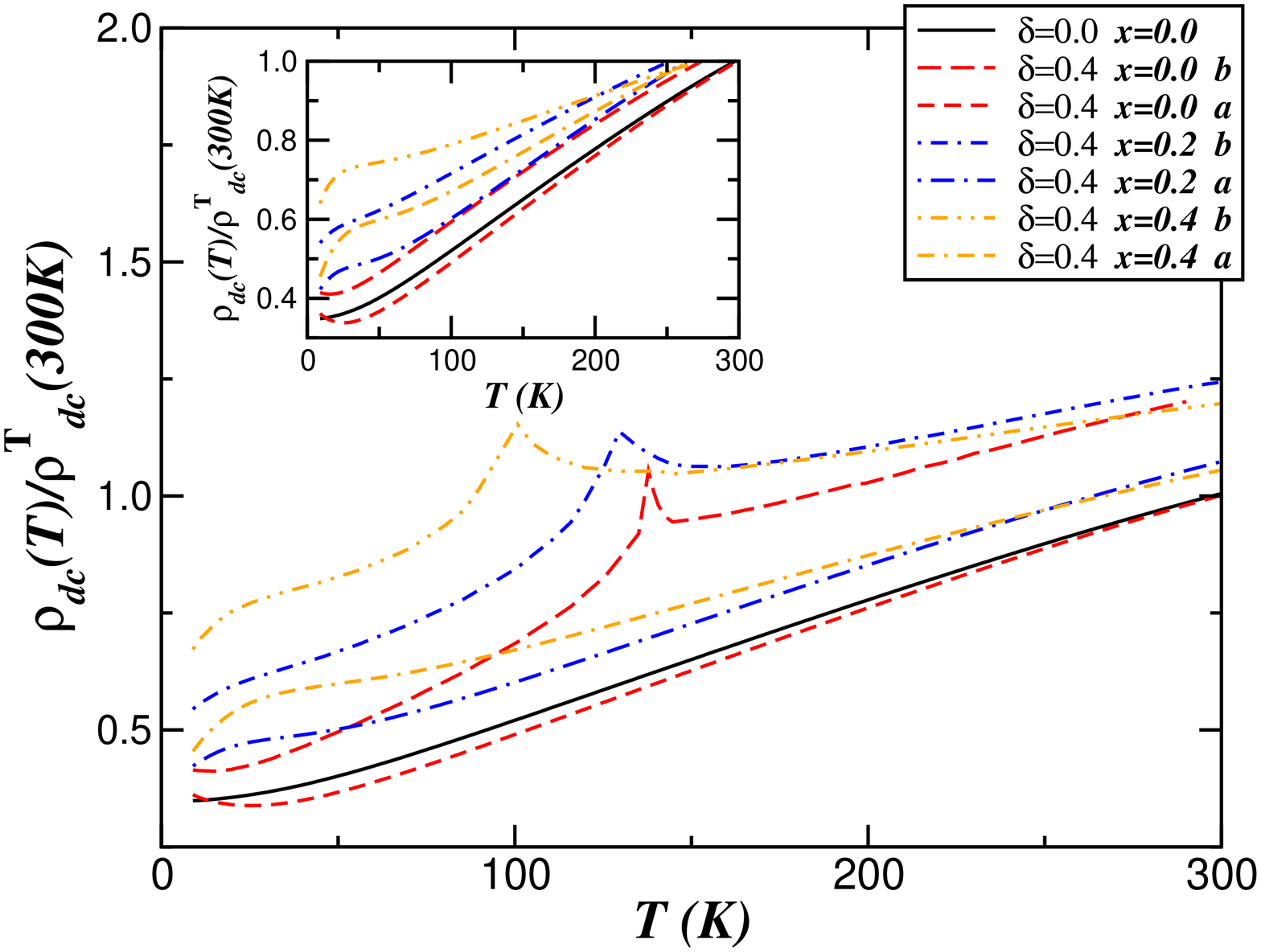}
\caption{(Color online)
Resistivity versus temperature, showing the in-plane resistivity 
anisotropy with (main panel) and without (inset) the non-local, static 
(Ising) 
nematic correlations. Notice that DMFT already yields the correct sign 
of the anisotropy, and non-local corrections further enhance it near 
$T_{T-O}$, in good accord with data.}
\label{fig3}
\end{figure}

{\it If} incipient nematicity exists above $T_{T-O}$, one expects  
electronic anisotropy to persist up to a sizable $T>T_{T-O}$: this should 
manifest as a in-plane resistivity anisotropy above $T_{T-O}$.  In addition 
to the DMFT contributions, we also accounted for the influence of the 
(short-range) non-local ON (plus T-O) fluctuations on the $dc$ resistivity 
(these are {\it necessary} close to $T_{T-O}$) by including coupling of 
carriers to the {\it static}, $T$-dependent nematic susceptibility~\cite{sro} using an embedded cluster-variation method for $H_{eff}$ above. We used the 
DMFT effective masses and $J_{1},J_{2}$ as above in the evaluation of the 
short-range order contribution to $\rho_{a,b}(T)$. This nicely fits within 
our DMFT-based view, since non-local (Ising) correlation effects are only 
included at a static level.  In Fig.~\ref{fig3}, we show the total $dc$ 
resistivities $\rho_{a}(T)$ and $\rho_{b}(T)$ calculated within 
LDA+DMFT-plus static non-local ON correlations. Quite remarkably, we find 
that the in-plane 
resistivity anisotropy indeed survives up to $T\simeq 250-300$~K, sizably 
above $T_{s}=150$~K, though it is a bit overestimated at high $T$. It is 
instructive to notice that this anisotropy already exists within DMFT 
(inset to Fig.~\ref{fig3}), and is enhanced sizably near $T_{T-O}$ due to 
non-local nematic correlations (which peak at $T_{T-O}$ in our $H_{eff}$), 
in agreement with data. Here, $\rho_{b}(T)>\rho_{a}(T)$ in spite of 
$b<a~(\delta(T)>0)$ naturally follows from increased tendency to (Mott) 
localization in the $xz$ band sector (corresponding to orbitals pointing 
along $b$) across the ON-plus T-O transition (see Fig.~\ref{fig1}). Very 
satisfyingly, this anisotropy {\it increases} with $x$, as indeed  observed 
in experiment (in a 122-FePn recently, but this also could be seen 
in 1111-FePn systems, where $T_{T-O}$ and $T_{SDW}$ are separated). We 
emphasize that this is in conflict with pure $J_{1}-J_{2}$ {\it spin-only} 
models, where one expects $\rho_{b}(T)<\rho_{a}(T)$. It is also inconsistent 
with a weakly correlated Landau Fermi liquid (LFL) view for same reasons.  
In contrast, a ON-plus T-O transition within our OS-Mottness view fulfils 
{\it all} above experimental constraints~\cite{fisher}. Finally, the 
optical conductivity in DMFT (solely involving the full DMFT Green 
functions, $G_{xz}(\omega),G_{yz}(\omega)$), immediately implies 
different optical conductivities along $a,b$. Sizable SWT, $O(2.0)$~eV, 
across this ON-plus T-O transition in DMFT DOS above should show up as 
SWT in $\sigma_{a,b}(\omega)$ across $T_{T-O}$. Remarkably, 
{\it both} these features are indeed seen~\cite{fisher1}, lending 
strong support for our view.

A finite $\langle N\rangle >0$ also implies ferro-orbital (FO) ordering,
and, with sizable magnetoelastic coupling, leads to conditions propituous 
for the ${\bf Q}=(\pi,0)$ striped SDW order to occur. Microscopically, the
``localized'' orbital pseudospins directly follow from enhanced tendency to 
Mott localization of the $xz$ orbital states in our work. The 
``super-exchange'' couplings in the effective $J_{1}-J_{2}$ spin model, with
``localized'' spins arising from the incoherent Hubbard bands~\cite{si}, are 
now explicit functions of the neighboring (Ising) orbital correlations along 
$a$ and $b$~\cite{phillips}, and, for realistic $J_{1a},J_{1b},J_{2}$, the 
striped AF-SDW state is indeed the ground state. Thus, 
AF-SDW order will always follow, or be coincident with, the T-O distortion, 
as is well known. Interestingly, the anisotropy of the spin-wave 
excitations~\cite{analytis} in ${\bf q}$-space also follows from such a 
``localized'', frustrated model~\cite{si}.  A weak coupling LFL 
approach~\cite{mimin} can also achieve this upon ``tweaking'' the 
ellipticity of the electron pockets phenomenologically.  However, it 
is clearly inconsistent with $\rho_{b}(T) > \rho_{a}(T)$, as well as 
with the insulator-like $\rho_{b}(T)$ above $T_{T-O}$.  Finally, our 
work is not inconsistent with LDA+DMFT works for the AF-SDW 
state~\cite{valenti} where co-existing quasi-itinerant and localized 
band states are invoked; however, it bares the intimate connection 
between the ON-plus T-O transition and subsequent AF-SDW in FePn.
      
Thus, our work offers a unified rationalization of various unusual
features seen recently in underdoped 122-FePn. The observed in-plane 
resistive and optical anisotropies are understood as consequences of an 
{\it electronic} ON-plus orthorhombic order arising as a p-h instability 
(hence, {\it competing} with U-SC~\cite{sachdev}) of the ``high'' $T$ 
OS-incoherent metal.  Finding of nanoscale nematic domains in 122 systems, 
as well as nanoscale {\it electronic} order in the 1111 systems is not 
inconsistent with effects of (doping induced) static disorder on our 
ON-plus orthorhombic state, which falls into the ``liquid gas'' 
universality class of an Ising model in a (random in presence of disorder)
zeeman field~\cite{carlson}. To the extent that such fluctuating ON order 
will induce a spin gap and two distinct {\it local} ``environments'' 
(corresponding to $\langle N\rangle = +n,-n$), this should 
``rationalize'' NMR data~\cite{buechner} as well.  Our work opens up an 
intriguing possibility: an ON order with a T-O distortion, with subsequent 
stripe-AF, might be the {\it primary} competitor to unconventional SC in 
FePn, and establishes the important role of sizable MO 
correlations and OS Mott physics in this context.
    
L.C. thanks the Physical Chemistry departement at Technical University 
Dresden for hospitality.

\end{document}